\documentclass[proceedings, preprint]{rmaa}

\usepackage{paralist}

\usepackage{psfrag,color}

\SetYear{2011}
\SetConfTitle{Fourth science with the GTC meeting}

\title{First Science with SHARDS: emission line galaxies} 

\author{
  Antonio Cava,\altaffilmark{1} 
  Pablo G. P\'erez Gonz\'alez,\altaffilmark{1}
  and the SHARDS Team\altaffilmark{2}}

\altaffiltext{1}{Departamento de Astrof\'{\i}sica, Universidad Complutense de Madrid, E28040-Madrid, Spain (acava@fis.ucm.es).}

\altaffiltext{2}{http://guaix.fis.ucm.es/$\sim$pgperez/SHARDS/team.html}

\shortauthor{Cava et al.}
\shorttitle{First science with SHARDS: ELGs}

\listofauthors{A. Cava, P. G. P\'erez-Gonz\'alez, \& SHARDS collaboration}

\abstract{SHARDS is an unbiased ultra-deep spectro-photometric survey
  with GTC@OSIRIS aimed at selecting and studying massive passively
  evolving galaxies at z$=$1.0--2.3 using a set of 24 medium-band
  filters (FWHM$\sim$17~nm) at 500--950~nm in GOODS-N. Our observing
  strategy is optimized to detect at z$>$1 the prominent Mg absorption
  feature at rest-frame $\sim$280~nm, a distinctive, necessary, and
  sufficient feature of evolved stellar populations. Nonetheless, the
  data quality allow a plethora of studies on galaxy populations,
  including Emission Lines Galaxies (ELGs) about which we have started
  our first science verification project presented in this
  contribution.}

\resumen{SHARDS  (Survey for High-z Absorption Red \& Dead Sources) es una exploraci\'on espectro-fotom\'etrica
  ultra-profunda en GOODS-N que utiliza el instrumento OSIRIS@GTC para
  seleccionar galaxias masivas y pasivas a alto desplazamiento al rojo
  por medio de un conjunto de 24 filtros de anchura media en el rango
  \'optico. Nuestra estrategia de observaci\'on est\'a optimizada para
  detectar, en galaxias a z$>$1, la caracter\'istica banda de
  absorpci\'on del Mg(UV), que es capaz de identificar de forma
  un\'{\i}voca poblaciones estelares evolucionadas. Adem\'as, la
  calidad de los datos SHARDS permite enfrentarse a una multitud de
  diferentes estudios sobre evoluci\'on de galaxias, incluyendo las
  galaxias con l\'{\i}neas de emisi\'on que presentamos en esta
  contribuci\'on.}

\addkeyword{Galaxies: emission lines}
\addkeyword{Galaxies: evolution}
\addkeyword{Galaxies: general}
\addkeyword{Surveys}

\begin{document}
\maketitle

\section{Introduction}
\label{sec:intro}
The main driver of SHARDS is building and analyzing in detail an
unbiased sample of high-{\it z} quiescent galaxies, which extends to
fainter magnitudes ($\ge 26.5~$ABmag) the samples selected with color
techniques and spectroscopic surveys. This goal is of special
relevance for our understanding of galaxy formation and evolution.
Indeed, one of the most interesting results in Extragalactic Astronomy
in the last decade is the discovery of a numerous population of
massive galaxies ($\mathcal{M}$$>$10$^{11}$~$\mathcal{M}_\odot$) at
high redshift (Yan et al. 1999, Franx et al. 2003). These galaxies are
characterized by very small sizes, and thus large mass densities
(Trujillo et al.  2007). Some of them are already evolving passively
(Daddi et al.  2004), being good candidates for the progenitors of
massive nearby ellipticals (Hopkins et al. 2009). The existence of
very compact massive dead galaxies at high-redshift is extremely
challenging for models of galaxy formation, based on the hierarchical
$\Lambda$CDM paradigm (e.g.\,
de Lucia et al. 2006).
SHARDS will also provide reliable observational estimations of
relevant quantities such as galaxy number density, photometric
redshifts, stellar masses, ages, and sizes.
\begin{figure}[!b]
  \includegraphics[width=\columnwidth]{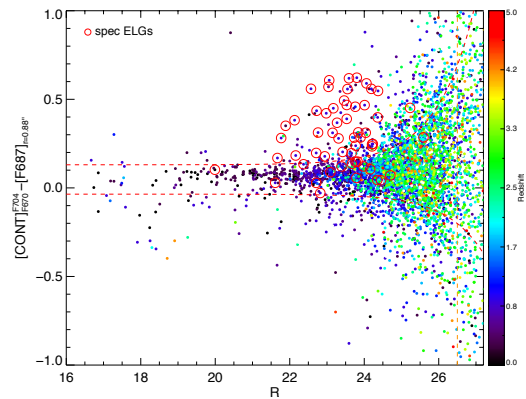}
  \caption{Illustration of an ELGs detection diagram for the filter
    F687W17.  Shaded lines show the 3-$\sigma$ threshold used to
    separate and define the locus occupied by ELGs.}
  \label{fig:simple}
\end{figure}
\begin{figure*}[!t]
\begin{centering}
  \includegraphics[width=15.9cm]{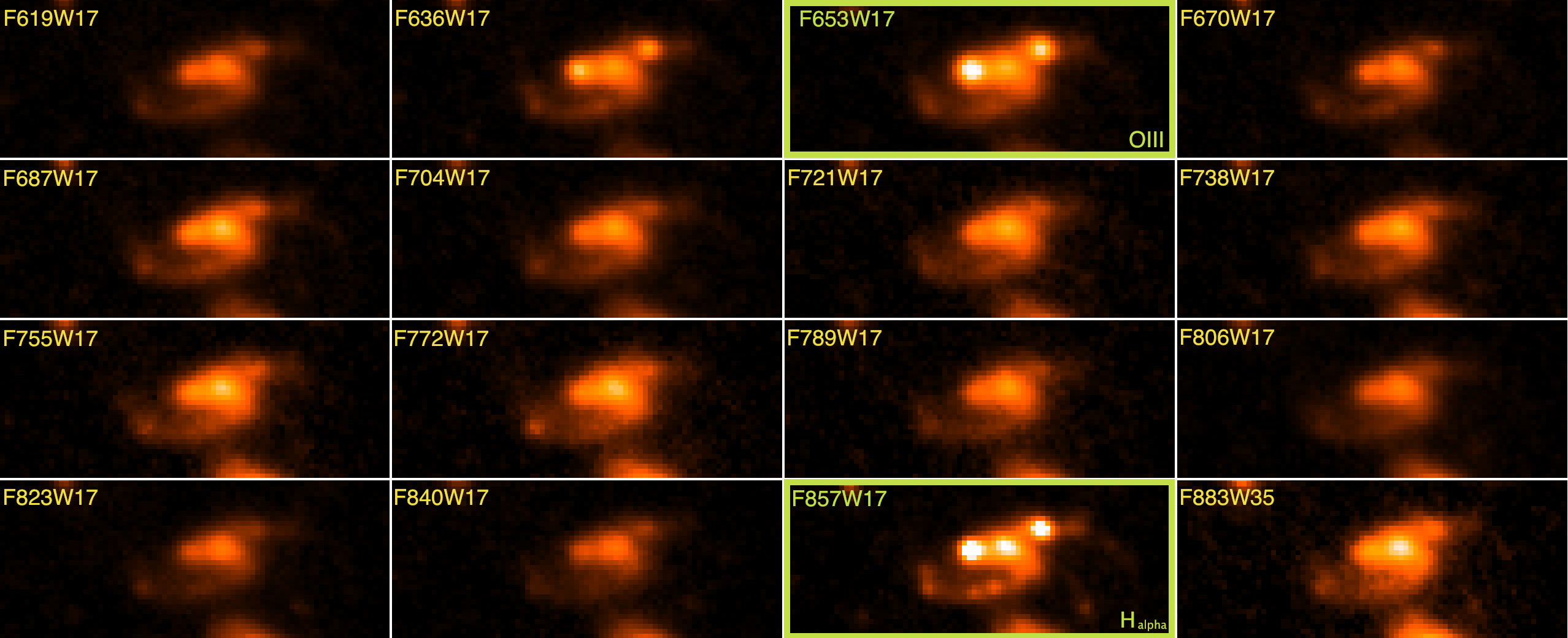}
  \caption{Example of emission lines detected on the SHARDS images of
    a nearby galaxy.  The mosaic shows the maps taken in the 16 SHARDS
    filters observed in 2011A. We clearly detect [OIII] in the F653W17
    filter and H$\alpha$ in the F857W17 filter for a galaxy at
    redshift $\sim$0.3.}
\end{centering}
  \label{fig:simple}
\end{figure*}
SHARDS is already in an advanced stage of progress regarding
observations and data reduction. Around $75\%$ of the planned
observations have already been carried out and a preliminary complete
data reduction is already available for first investigations.  The
program will be fully executed by the end of semester 2012A. A full
description of the Survey will be provided in P\'erez-Gonz\'alez et
al. (in prep., see also P\'erez-Gonz\'alez et
al. 2012 in this Volume). In addition to the main goal of the survey, a plethora
of secondary projects have been devised. In this GTC Science Meeting,
we presented the first analyses of emission line galaxies (ELGs) in
SHARDS.

\section{ELGs in SHARDS}

Following the approach adopted in previous works carried out by our
group at UCM (see Pascual et al. 2001, Villar et al.  2008, 2011), we
search for candidate ELGs using different selection diagrams based on
color-magnitude plots for each SHARDS filter.  The assumption is that
ELGs will present an excess of luminosity in a given filter with
respect to the continuum, estimated by using 2 contiguous filters, and
that a region with high probability to find ELGs can be statistically
defined by a sigma-clipping procedure. An illustration is given in
Fig.1 for the filter F687W17 (central wavelength at 687~nm and width
17~nm) where we can clearly see how the method is able to separate
most of the ELGs from the whole galaxy population.  In fact, by
definition, all the galaxies above the upper envelope are potentially
ELG candidates. The effectiveness of the method is confirmed by
plotting the spectroscopically confirmed ELGs (red open circles depict
[OII] emitters) showing that the vast majority of them ($\sim 75\%$) fall in the
expected ELG region above the upper dashed line. Using this technique,
a very large sample of ELGs at different redshifts can be selected.
Having all the photometric redshift in hand (P\'erez-Gonz\'alez et al.
2008) we will also be able to disentangle between different detected
emission lines. We are currently calibrating our method comparing with
spectroscopic observations from literature in order to provide
reliable equivalent widths, line fluxes and emission line based star
formation rate (SFR) that will allow us a comparison of different SFR
indicators derived from the UV, optical and MIR/FIR. We expect this
work will provide the largest sample of ELGs in GOODS-N available up
to date for evolutionary studies. As an additional check, we show in
Fig.2 an example of a mosaic built using all the observed filters for
one of the galaxies selected in the ELG region.  The mosaic shows a
clear brightness excess in the SHARDS filters corresponding to [OIII]
and H${\alpha}$ emissions at the corresponding redshift ($z\sim0.3$ in
this case) for the selected galaxy. The depth and the quality of
OSIRIS/GTC imaging allow to study in detail the region where the
emission is originated, providing additional information on the
possible involved physical processes. We will address in detail this
topic in a forthcoming paper (Cava et al., in prep.) but we stress the
encouraging first results we are already obtaining despite the
preliminary reduction and calibration of the data.

\end{document}